\documentclass[twocolumn,PRB,aps,psfig,preprintnumbers,superscriptaddress]{revtex4}
\usepackage{graphicx}
\usepackage{epstopdf}
\usepackage{float}
\usepackage{color}
\usepackage{amsmath}

\newcommand{\red}[1]{\protect\textcolor{red}{#1}}

\begin{document}

\title{Itinerant G-type antiferromagnet SrCr$_2$As$_2$ studied by magnetization, heat capacity, electrical resistivity, and NMR measurements}
\author{Q.-P. Ding}
\affiliation{Ames Laboratory, Iowa State University, Ames, Iowa 50011, USA}
\affiliation{Department of Physics and Astronomy, Iowa State University, Ames, Iowa 50011, USA}
\author{Santanu Pakhira}
\affiliation{Ames Laboratory, Iowa State University, Ames, Iowa 50011, USA}
\author{N. S. Sangeetha$\footnote[1]{Present address: Institute for Experimental Physics~IV, Ruhr University Bochum, 44801 Bochum, Germany}$}
\affiliation{Ames Laboratory, Iowa State University, Ames, Iowa 50011, USA}
\author{E. H. Krenkel}
\affiliation{Ames Laboratory, Iowa State University, Ames, Iowa 50011, USA}
\affiliation{Department of Physics and Astronomy, Iowa State University, Ames, Iowa 50011, USA}
\author{E. I. Timmons}
\affiliation{Ames Laboratory, Iowa State University, Ames, Iowa 50011, USA}
\affiliation{Department of Physics and Astronomy, Iowa State University, Ames, Iowa 50011, USA}
\author{M. A. Tanatar}
\affiliation{Ames Laboratory, Iowa State University, Ames, Iowa 50011, USA}
\affiliation{Department of Physics and Astronomy, Iowa State University, Ames, Iowa 50011, USA}
\author{R. Prozorov}
\affiliation{Ames Laboratory, Iowa State University, Ames, Iowa 50011, USA}
\affiliation{Department of Physics and Astronomy, Iowa State University, Ames, Iowa 50011, USA}
\author{D. C. Johnston}
\affiliation{Ames Laboratory, Iowa State University, Ames, Iowa 50011, USA}
\affiliation{Department of Physics and Astronomy, Iowa State University, Ames, Iowa 50011, USA}
\author{Y. Furukawa}
\affiliation{Ames Laboratory, Iowa State University, Ames, Iowa 50011, USA}
\affiliation{Department of Physics and Astronomy, Iowa State University, Ames, Iowa 50011, USA}

\date{\today}

\begin{abstract} 
   The physical properties of itinerant antiferromagnetic (AFM) SrCr$_2$As$_2$ with body-centered tetragonal ThCr$_2$Si$_2$ structure were investigated in single crystalline  and polycrystalline forms by electrical resistivity $\rho$, heat capacity $C_{\rm p}$, magnetic susceptibility $\chi$ versus temperature~$T$ and magnetization $M$ versus applied magnetic field $H$ isotherm measurements as well as  $^{75}$As and $^{53}$Cr nuclear magnetic resonance (NMR) measurements in the  wide temperature range $T$ = 1.6--900 K. 
   From the $\chi(T)$ and $^{75}$As NMR measurements, the G-type AFM state below $T_{\rm N}$ = 615(15) K has been determined, consistent with  the previous neutron-diffraction measurements. 
    Direct evidence of magnetic ordering of the Cr spins was shown by the observation of the $^{53}$Cr NMR spectrum under $H$ = 0. 
    From the $\chi(T)$ measurements on single-crystal SrCr$_2$As$_2$  under the two different magnetic field directions $H$ $||$ $ab$ and  $H$ $||$ $c$ in the AFM state,  the Cr ordered moments are shown to align along the $c$ axis in the G-type AFM state.   
 % and linear behavior above it.
 %    This observation bears some similarity to the parent compounds of iron-based superconductors BaFe2As2 and SrFe2As2, but not CaFe2As2 \cite{}.  
  % and the ordered Cr magnetic moment $\langle \mu \rangle$ is estimated to be $\sim$ 2.36 $\mu_{\rm B}$ at $T$ = 1.6 K .
      The metallic state is directly evidenced by the  $\rho$, $C_{\rm p}$, and NMR measurements, and the density of states at the Fermi energy  ${\cal D}(E_{\rm F})$ in the AFM state is estimated to be 7.53 states/eV f.u. for both spin directions which is almost twice the bare ${\cal D}(E_{\rm F})$ estimated from first-principles calculations, suggesting an enhancement of the conduction-carrier mass by a factor of two in the AFM state. 
The ${\cal D}(E_{\rm F})$  is found to be nearly constant below at least  100 K and is independent of $H$.  
   The $\rho(T)$ is found to show $T$-linear behavior above $T_{\rm N}$ and exhibits  positive curvature below $T_{\rm N}$ where significant loss of spin-disorder scattering upon magnetic ordering is observed. 
   The resistivity anisotropy of the compound remains moderate $\rho_c/\rho_a \sim$9 through most of the magnetically-ordered phase but shows a rapid increase below 50~K. 

  %    The magnetic measurements also confirm that the ordered Cr moments are itinerant rather than localized.

\end{abstract}

%  \pacs{74.70.Xa, 76.60.-k}
\maketitle

 \section{Introduction}
   Since the discovery of high $T_{\rm c}$ superconductivity in iron pnictides \cite{Kamihara2008}, the interplay between electron correlations  and unconventional superconductivity (SC) has attracted much interest. 
    In most of the Fe-based  superconductors (FBSC), the parent materials are metallic and exhibit antiferromagnetic (AFM) ordering below the N\'eel temperature $T_{\rm N}$ \cite{Canfield2010,Johnston2010,Stewart2011}.
    SC in these compounds emerges upon suppression of the AFM phase by the application of pressure and/or chemical substitution. 
  Because of  the proximity between the magnetically-ordered and SC phases,  the effects of electron correlations on the appearance of SC are naturally considered to be important.
%  The electron correlations could be caused by strong Hund\rq{}s exchange interactions for $3d$ electrons depending on the number of $d$ electrons. 

   The role of the electron correlations can be investigated by varying the 3$d$ transition-metal elements since electron correlations could be caused by strong Hund\rq{}s exchange interactions for $3d$ electrons. % depending on the number of $d$ electrons.
   In the case of $3d^5$ configurations, the half-filled $3d$ electron shells  will lead to a Mott insulating state. 
   For example, an insulating state is observed in BaMn$_2$As$_2$ [Mn$^{2+}$ ($S$ = 5/2, 3$d^5$)] which exhibits an AFM state below a high N\'eel temperature $T_{\rm N}$ = 625 K \cite{Singh2009_Mn,Johnston2011}. 
    For FBSCs such as $A$Fe$_2$As$_2$ ($A$ = Sr, Ba, Ca), a nearly 3$d^6$ electron configuration ($S$~=~2)  is expected for the Fe ions in the metallic state.
    On the other hand, 3$d^4$  electron configurations are expected for Cr$^{2+}$ ions in Cr-based compounds such as $A$$^{2+}$Cr$_2$As$_2$.
    This is an interesting system since  a $3d^4$ electron configuration ($S$ = 2) can be considered as a mirror of the 3$d^6$ electron configuration for Fe$^{2+}$ ions  in the  parent materials of FBSC such as $A$$^{2+}$Fe$_2$As$_2$ with respect to the 3$d^5$ electron configuration.

\begin{figure}[tb]
\includegraphics[width=\columnwidth]{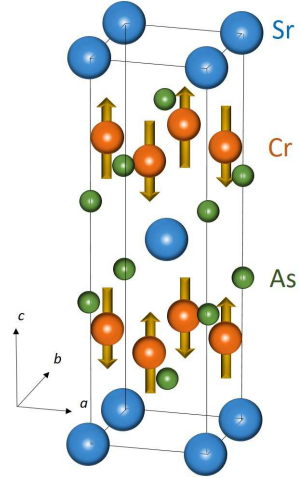} 
\caption{Crystal and G-type magnetic structures of SrCr$_2$As$_2$. The arrows represent the direction of the Cr ordered moments in the antiferromagnetic state. After Ref.~\cite{Das2017}.  
 }
\label{fig:structure}
\end{figure}   

   Recent theoretical proposals for the possible appearance of superconductivity in Cr-based compounds such as  electron-doped LaOCrAs \cite{Wang2017,Pizarro2017} and negative-pressurized and/or electron-doped BaCr$_2$As$_2$ \cite{Edelmann2017} as well as the discovery of SC in CrAs under pressure \cite{Kotegawa2014,Wu2014}, and in $A_2$Cr$_3$As$_3$ ($A$ = Na, K, Rb, Cs) \cite{Bao2015,Tang2015,Tang2015_2,Mu2018} and $A$Cr$_3$As$_3$ ($A$ = K, Rb) \cite{Tang2015_3} also lead to further interest in investigating Cr-based compounds.
  BaCr$_2$As$_2$,  one of the $A$Cr$_2$As$_2$ systems with the body-centered tetragonal ThCr$_2$Si$_2$-type  structure, is a metallic G-type antiferromagnet with a N\'eel temperature $T_{\rm N} \sim$ 580(10) K with a Cr ordered moment of 1.9 $\mu_{\rm B}$/Cr aligned along the $c$ axis  \cite{Filsinger2017}.
%where Cr$^{2+}$ moments (1.9 $\mu_{\rm B}$) align along the $c$  \cite{Filsinger2017}. 
    A reduction in electron correlations in BaCr$_2$As$_2$ in comparison with the half-filled 3$d^5$ case has been reported from angle-resolved photoemission spectroscopy (ARPES) measurements \cite{Nayak2017, Richard2017},  where a stronger Cr-As covalency relative to the Fe-based SCs  was suggested to be an important factor in the appearance of SC. 
      The isostructural compound SrCr$_2$As$_2$ has been reported to exhibit the same G-type AFM ordering with Cr ordered moments 1.9 $\mu_{\rm B}$/Cr (see Fig. \ref{fig:structure}) below $T_{\rm N}$ = 590(5)~K from neutron diffraction (ND) and magnetic susceptibility versus temperature  $\chi(T)$ measurements on polycrystalline samples \cite{Das2017}. 
Although the reduction of the Cr ordered moments of 1.9 $\mu_{\rm B}$/Cr from 4 $\mu_{\rm B}$/Cr expected for a localized state of Cr$^{2+}$ ion  in SrCr$_2$As$_2$ suggests an itinerant nature \cite{Das2017}, there is no direct experimental evidence of a metallic state in the compound. 

    No electrical resistivity $\rho(T)$ study has been reported on  SrCr$_2$As$_2$ in either polycrystalline or single crystalline forms. 
Even for the isostractrual compound BaCr$_2$As$_2$, little is known about the behavior of $\rho(T)$.  
   In-plane resistivity measurements were made up to 300~K in BaCr$_2$As$_2$ \cite{Filsinger2017, Richard2017}, and found a notably non linear temperature dependence. 
     Because of the high magnetic ordering temperature $\sim$600~K, these measurements merely cover half of $T_{\rm N}$. 
   The resistivity change through the magnetic transition provides an important insight into the changes of the Fermi surface due to magnetic superzone boundaries and the contribution of magnetic scattering, so getting this information is important. 
%    In particular,  no resistivity study was done on  SrCr$_2$As$_2$ in either polycrystalline or single crystalline form.  
    In addition, there is no information on the anisotropy of $\rho(T)$ in these compounds. 
   Therefore it is important  to obtain the information of not only in-plane but also inter-pane $\rho(T)$ up to higher temperature above $T_{\rm N}$  in SrCr$_2$As$_2$.     
Furthermore, no measurements such as heat capacity $C_{\rm p}$ and magnetization $M$ versus applied magnetic field $H$ isotherm measurements 
to characterize the physical properties of  SrCr$_2$As$_2$ have been reported so far, although recently a Raman spectroscopy study on  single crystalline  SrCr$_2$As$_2$ has been performed \cite{Kaneko2021}.

   In this paper, we report a comprehensive study of both polycrystalline and single crystalline SrCr$_2$As$_2$ in a wide temperature range from 1.6 K to 900 K from $\rho$, $M$, $C_{\rm p}$ and nuclear magnetic resonance (NMR) measurements.
 After the description of the experimental details in Sec.~II, we first show the results of x-ray diffraction measurements on a polycrystalline sample in Sec.~III.~A. 
Then the results of $\chi(T)$ and $M(H)$ measurements are reported in Sec.~III.~B.
In Sec.~III.~C,   the $C_{\rm p}(T)$ data are presented. 
% and $\rho(T)$ 
The results of the in-planer and inter-plane electrical resistivities of SrCr$_2$As$_2$ are discussed in Sec.~III.~D.  
The results of $^{75}$As and $^{53}$Cr NMR measurements are shown in Sec.~III.~E.  
Finally summary of our results is given in Sec.~IV.

%    The G-type spin structure has been evidenced by $^{75}$As NMR spectra in the AFM state.
%    The temperature ($T$) denpendence of the linewidth of $^{75}$As NMR spectra may suggest the three-dimensional nature of the magnetism.
 %   The nuclear spin-lattice relaxation rate 1/$^{75}T_1$ data show negligible anisotropy in SrCr$_2$As$_2$.
 %    The ordered Cr magnetic moment $\langle \mu \rangle$ $\sim$ 2.36 $\mu_{\rm B}$ can be estimated from zero-field $^{53}$Cr NMR spectra at $T$ = 1.6 K.  

%  After the description of the experimental details in Sec. II, we first show the results of x ray diffraction measurements on  polycrystalline sample in Sec. III. A. Then , the results of  magnetic susceptibility and isothermal magnetization versus measurements are reported in Sec. III. D. In Sections III. C.and D,   the specific heat data and resistivity data are  described, respectively.  The the results of $^{75}$AS and $^{53}$Cr NMR measurements are shown in Sec. III. E.   Finally we summarize our results in Sec. IV.

 \section{Experimental details}

A polycrystalline sample of SrCr$_2$As$_2$ was prepared by conventional solid-state reaction using high-purity Sr (Sigma Aldrich, 99.95\%), Cr (Alfa Aesar, 99.99\%) and As (Alfa Aesar, 99.999 99\%). Dendritic Sr pieces were taken with prereacted CrAs in the molar ratio Sr:CrAs~=~1.05:2. The additional amount of Sr was to compensate for the loss of Sr due to evaporation and to reduce the presence of CrAs impurity phase. The Sr and CrAs mixture was pelletized and loaded into an alumina crucible followed by sealing in an evacuated silica tube under $\approx$ 1/4 atm of high-purity Ar gas. The assembly was then heated to 900~$^{\circ}$C  at the rate of 100~$^{\circ}$C/h and held there for 48~h followed by furnace-cooling to room temperature. The heat-treated pellet was then reground inside a He-filled glove box and pelletized again for an additional heat treatment as just described. In order to ensure homogeneous phase formation, this process was repeated once more with intermediate grinding. Finally, the sample was heated to 1150~$^{\circ}$C at a rate of 100~$^{\circ}$C/h and annealed there for 48 h followed by furnace-cooling.

Room-temperature powder x-ray diffraction (XRD) measurements on the powdered polycrystalline sample were carried out using a Rigaku Geigerflex x-ray diffractometer with Cu-$K_\alpha$ radiation. Structural analysis was performed by Rietveld refinement using the FULLPROF software package~\cite{Carvajal1993}.

Single crystals of SrCr$_2$As$_2$ were grown by the solution-growth technique using Sn flux with a molar ratio Sr:Cr:As:Sn = 1.1:2:2:10. The starting elements described above were loaded into an alumina crucible which was then sealed in a silica tube under $\approx$ 1/4 atm of high-purity Ar gas. The sealed tube was then heated to 600~$^{\circ}$C at a rate of 50~$^{\circ}$C/h where the temperature was held for 6~h. The temperature was then ramped to 1100~$^{\circ}$C/h at a rate of 100~$^{\circ}$C/h and held there for 20 h. The assembly was then cooled to 500~$^{\circ}$C at a rate of 3~$^{\circ}$C/h where it was centrifuged to remove the crystals from the remaining flux. Shiny crystal plates with typical dimensions $2 \times 2 \times 0.2$~mm$^3$ were obtained.

The chemical composition and homogeneity of the crystals were checked using a JEOL scanning-electron microscope (SEM) equipped with an energy-dispersive x-ray spectroscopy (EDS) analyzer.  $M$ measurements were carried out using a Magnetic Property Measurement System (MPMS) from  Quantum Design, Inc., in the $T$ range 1.8--300 K and with $H$ up to 5.5~T (1~T~$\equiv10^4$~Oe). 
 The high-temperature $M$ measurements from 300 to 900 K were performed using the vibrating-sample magnetometer (VSM) option for a Physical Property Measurement System (PPMS, Quantum Design, Inc.).
The $C_{\rm p}$ and four-probe $\rho$ measurements were carried out using the PPMS. The $C_{\rm p}(H,T)$ was measured using a relaxation technique\@.

   Samples for resistivity study  were prepared from thin slabs, cleaved out of the crystal with two fresh surfaces corresponding to the tetragonal $ab$ plane.
   For four-probe in-plane resistivity measurements, the bar-shaped samples were cleaved with the long side corresponding to the tetragonal $a$ axis with typical dimensions 2$\times$0.3$\times$0.05~mm$^3$. 
    For measurements  in the range 1.8 to 400~K, contacts to the samples were prepared by soldering 50~$\mu$m diameter silver wires using tin as solder, a technique also used for creating contacts to iron-based superconductors \cite{anisotropy,SUST,patent}. 
   The contact resistance for SrCr$_2$As$_2$-Sn contacts turned out to be similarly low, several $\mu \Omega$, enabling two-probe interplane $c$-axis  resistivity measurements (see Refs.~\cite{anisotropy,anisotropy2} for details).
   The samples for interplane resistivity measurements were plate-shaped (typically 0.05~mm thick with 0.5$\times$0.5 mm$^2$ area) and their top and bottom surfaces were covered with Sn and two silver wires attached acting as current and voltage probes in the four-probe resistivity measurements. 
   The actual resistance measured in this experiment is a sum of sample and contact resistances, with the contact resistance representing approximately 1\% of the measured resistance. 
    Measurements in samples with the length along the current path significantly smaller than the cross-sectional dimensions suffer grossly from internal sample connectivity. 
    Because of this, measurements were performed on more than 20 samples to verify reproducibility.
  %  Four-probe resistivity measurements in the range 1.8 to 400~K were made in the PPMS.

   Whereas the PPMS measurements were performed in the temperature range 1.8 to 400 K,  high-temperature resistivity measurements up to 700~K  were performed in a home-made fixture placed in a vacuum furnace. 
   A platinum resistance thermometer PT-100 was used for temperature readings.  
    Contacts to the sample were made with conducting silver paste, with typical contact resistances in the 1 $\Omega$ range. 
    Four-probe resistivity measurements were made using an LS372 resistance bridge. 
    For the sake of comparison we measured the resistivity of a nickel wire in the same home-made furnace setup. 
    The wire was of 99.95\% purity from Alfa Aesar. 
   We found reasonable coincidence of the ferromagnetic transition temperature with the literature data \cite{Ni}, confirming the accuracy of the temperature readings (see the left inset in Fig.~8 below).

   NMR measurements of $^{75}$As ($I$ = $\frac{3}{2}$, $\frac{\gamma_{\rm N}}{2\pi}$~=~7.2919~MHz/T, $Q=$~0.29~barns) and $^{53}$Cr ($I$ = $\frac{3}{2}$, $\frac{\gamma_{\rm N}}{2\pi}$ = 2.40664~MHz/T, $Q=$~0.03~barns) nuclei were conducted using a lab-built phase-coherent spin-echo pulse spectrometer in a wide temperature range 1.6--630 K. 
    For $T \geq$ 300 K, $^{75}$As NMR spectra were measured in steps of resonance frequency $f$ by measuring the intensity of the Hahn spin echo at $H$ = 7.4089 T. 
    For $T \leq$ 300 K, $^{75}$As NMR spectra were obtained by sweeping $H$ at a fixed frequency $f$ = 51.1~MHz. 
     The $^{53}$Cr NMR spectrum at $T$ = 1.6 K was measured in steps of $f$ by measuring the intensity of the Hahn spin echo at $H$ = 0. 
   The $^{75}$As nuclear spin-lattice relaxation rate 1/$T_ 1$ was measured with a saturation-recovery method.
   $1/T_1$ at each $T$ was determined by fitting the nuclear magnetization $M$ versus time $t$ using the exponential function $1-M(t)/M(\infty) = 0.1 e^ {-t/T_{1}} +0.9e^ {-6t/T_{1}}$, where $M(t)$ and $M(\infty)$ are the nuclear magnetization at time $t$ after saturation and the equilibrium nuclear magnetization at $t$ $\rightarrow$ $\infty$, respectively.

\begin{figure}
\includegraphics[width=3.3in]{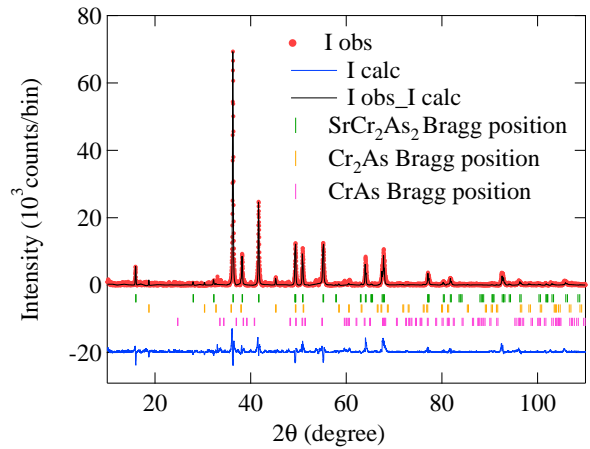}
\caption{X-ray diffraction pattern for polycrystalline SrCr$_2$As$_2$ obtained using a Rigaku powder diffractometer and Cu-$K_\alpha$ radiation. 
The observed data are shown by filled red circles, the allowed Bragg diffraction positions as vertical green bars, 
Cr$_2$As impurity peaks as vertical orange bars,  CrAs impurity peaks as vertical pink bars, the three-phase fit by the black curve, and the difference between the observed data and the two-phase refinement as the bottom blue trace. 
 }
\label{SrCr2As2_XRD}
\end{figure}

 \section{Results and discussion}

\subsection{Structural characterization}

The room-temperature XRD pattern of the polycrystalline SrCr$_2$As$_2$ sample is shown in Fig.~\ref{SrCr2As2_XRD}. The intense peaks could be indexed with the ThCr$_2$Si$_2$-type crystal structure with lattice parameters $a = b = 3.87(4)$\AA\ and $c = 12.89(2)$~\AA\@. For comparison, previously-reported room-temperature lattice parameters are $a=3.918(3)$~\AA\ and $c=13.05(1)$~\AA~\cite{Pfisterer1980}. 
The additional contributions to the x-ray diffraction pattern from minor amounts of CrAs (1.0 wt\%) and Cr$_2$As (5.3 wt\%) are shown by the three-phase refinement in Fig. 2.
 However, the magnetic contributions from these two impurity  phases are negligible and do not alter the observed intrinsic magnetic behavior of polycrystalline SrCr$_2$As$_2$ as reported earlier using neutron diffraction measurements~\cite{Das2017}. The SEM-EDS measurements of the single crystals revealed good homogeneity with the average composition SrCr$_{2.03(2)}$As$_{2.04(4)}$.

\subsection{\label{Sec:Magsuscep} Magnetic susceptibility and isothermal magnetization versus magnetic field measurements}

\begin{figure}
\includegraphics[width=3.3in]{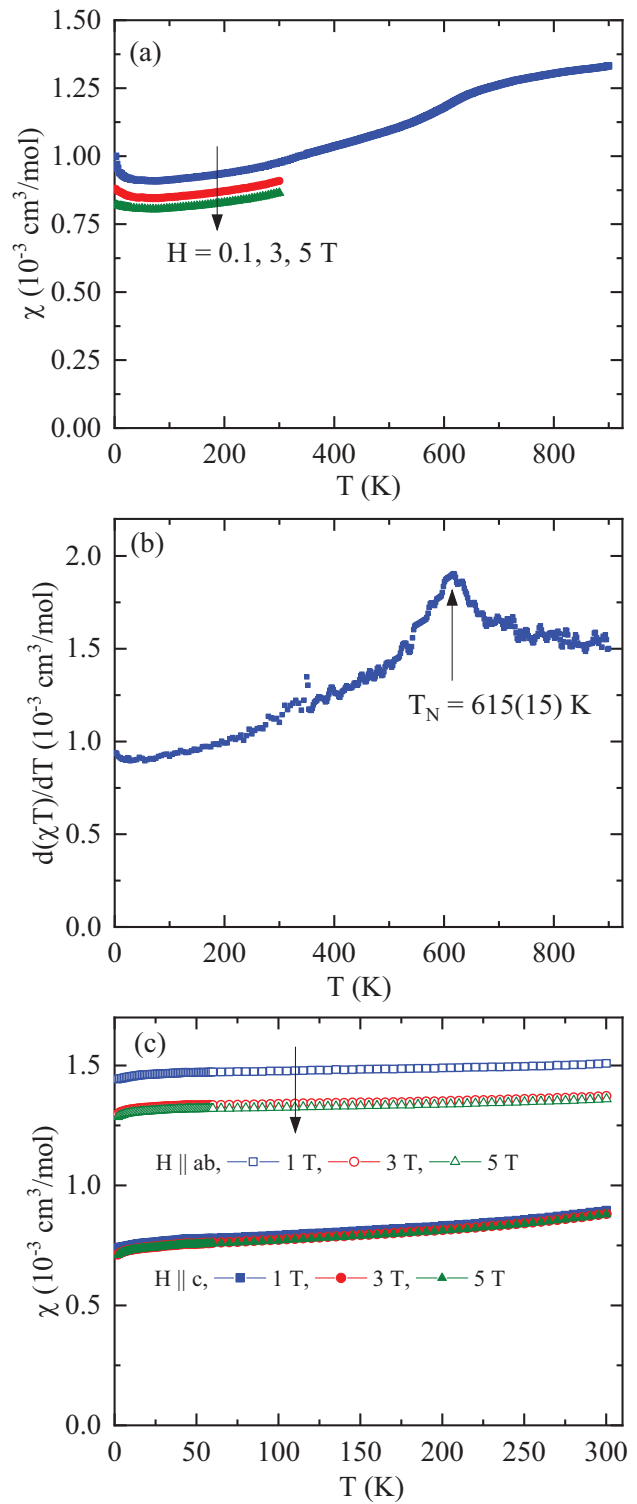}
\caption{(a) Temperature~$T$  dependence of the magnetic susceptibility $\chi = M/H$ for polycrystalline SrCr$_2$As$_2$ measured at different magnetic fields~$H$\@.  The low-$T$ upturns are likely associated with minor amounts of paramagnetic impurities that are suppressed at the higher fields. (b)~$d(\chi T)/dT$ vs~$T$ obtained using the data for $H=0.1$~T in~(a).  The peak in the plot gives the N\'eel temperature~\cite{Fisher1962} $T_{\rm N} = 615(15)$~K as indicated by an arrow.  (c) Anisotropy in $\chi(T)$ for a SrCr$_2$As$_2$ crystal measured at different fields for both $H \parallel ab$ and $H \parallel c$.  }
\label{Fig_M-T}
\end{figure}

The $T$ dependence of the zero-field-cooled (ZFC) magnetic susceptibility $\chi = M/H$ for polycrystalline SrCr$_2$As$_2$ measured in different magnetic fields is shown in Fig.~\ref{Fig_M-T}(a). There is no evidence for the first-order antiferromagnetic/paramagnetic transition in the CrAs impurity phase~\cite{Wu2010, Zhu2016} at 258~K\@.  In addition, the direction-averaged $\chi$ of CrAs is only $\approx 5.5$ and~$7.4\times 10^{-4}\,{\rm cm^3/mol}$ for $T\to 0$ and 300~K, respectively~\cite{Zhu2016}, which when multiplied by the small CrAs fraction of the sample is small compared to the measured polycrystalline data in Fig.~\ref{Fig_M-T}(a).   Additionally, the $\chi(T)$ data exhibit a small upturn below $\approx 35$~K that is likely due to the contribution of paramagnetic (PM) impurities. This contribution tends to saturate at high~$H$ as expected.

A plot of $d(\chi T)/dT$ versus~$T$ obtained from the data for $H=0.1$~T in~Fig.~\ref{Fig_M-T}(a) is shown in Fig.~\ref{Fig_M-T}(b).  The peak in this plot (the Fisher relation~\cite{Fisher1962}) gives the N\'eel temperature $T_{\rm N} = 615(15)$~K\@.  This value agrees with $T_{\rm N} = 590(5)$~K obtained in previous ND measurements of ${\rm SrCr_2As_2}$ which also showed that the ordering of the itinerant Cr moments is G-type with the moments aligned along the tetragonal $c$~axis~\cite{Das2017}.  The broad increase in $\chi(T)$ for $T > T_{\rm N}$ in Fig.~\ref{Fig_M-T}(a) is characteristic of two-dimensional AFM correlations persisting well above $T_{\rm N}$ in this itinerant antiferromagnet, similar to previous reports for the 122-type local-moment pnictide insulators SrMn$_2$As$_2$, CaMn$_2$As$_2$, SrMn$_2$Sb$_2$, SrMn$_2$Sb$_2$, CaMn$_2$P$_2$, and SrMn$_2$P$_2$~\cite{Sangeetha_2016, Sangeetha_2018, Sangeetha_2021}.

Figure~\ref{Fig_M-T}(c) shows the ZFC $\chi(T)$ data of a SrCr$_2$As$_2$ single crystal in the $T$ range 2--300~K measured at different fields applied in the $ab$~plane ($\chi_{ab}$) and along the $c$ axis ($\chi_{c}$). The $\chi(T)$ data are strongly anisotropic with $\chi_{ab} > \chi_{c}$ and $\chi_{ab}$ shows a weak $T$ dependence, as expected for a collinear antiferromagnet with $c$-axis moment alignment. The field-dependent anisotropy of $\chi_{ab}$ is likely due to a trace amount of ferromagnetic (FM) or PM impurity as discussed below. The $\chi_{c}(T)$ is also discussed below.

\begin{figure}
\includegraphics[width=3.3in]{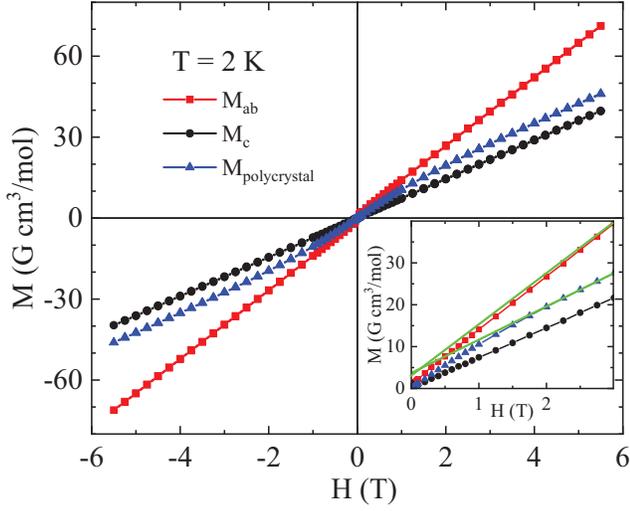}
\caption{Four-quadrant $M(H)$ hysteresis loops measured at $T=2$~K for single crystalline SrCr$_2$As$_2$ in two field directions and for polycrystalline SrCr$_2$As$_2$. Inset: Expanded low-$H$ region where the same weak nonlinearity is observed for the polycrystalline sample and in $M_{ab}(H)$ for the single crystal. Extrapolation of the higher-field data to $H=0$ (solid green lines) gives a FM or PM impurity magnetization of ${\sim \rm 3~G\,cm^3/mol=0.0005~\mu_{\rm B}}$/f.u.\ for both the polycrystalline and single-crystal samples.  
}
\label{Fig_Hysteresis_2K}
\end{figure}

 $M(H)$ hysteresis loops measured at $T = 2$~K for polycrystalline and single-crystal SrCr$_2$As$_2$ are shown in Fig.~\ref{Fig_Hysteresis_2K}. No magnetic hysteresis is found in either form of the material for either field direction. However, a weak negative curvature in the low-field $M(H)$ data for the polycrystalline sample and the $M_{ab}(H)$ data for the single crystal is observed (inset of Fig.~\ref{Fig_Hysteresis_2K}), which is likely associated with the contribution of FM or of PM impurities detected in the above $\chi(T)$ measurements.  At higher fields this contribution saturates. The saturation magnetization of the impurities is estimated by extrapolating the linear high-field $M(H)$ data to $H = 0$ and is found to be only $\approx 3$~G\,cm$^3$/mol~$\approx 0.0005\,\mu_{\rm B}$/f.u., where $\mu_{\rm B}$ is the Bohr magneton and f.u.\ means formula unit.

\begin{figure}
\includegraphics[width=3.3in]{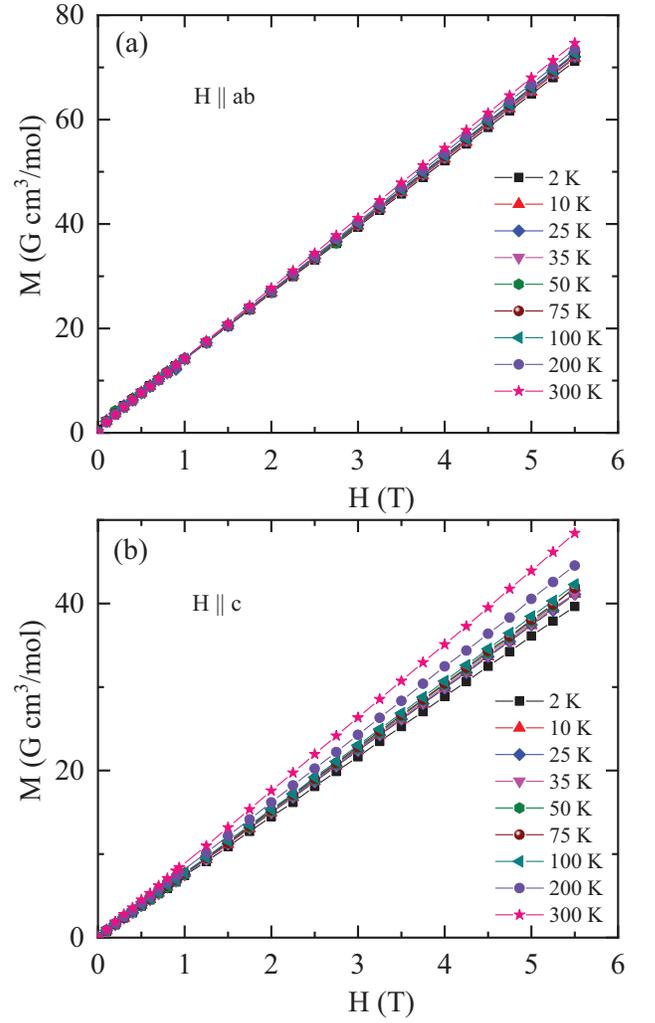}
\caption{$M(H)$ isotherms for a SrCr$_2$As$_2$ single crystal at different temperatures for (a) $H \parallel ab$ and (b) $H \parallel c$.}
\label{Fig_M-H}
\end{figure}

Figures~\ref{Fig_M-H}(a) and ~\ref{Fig_M-H}(b) show $M(H)$ isotherms of single-crystal SrCr$_2$As$_2$ measured at different temperatures for $H \parallel ab$ and $H \parallel c$, respectively. The $M(H)$ data for $H \parallel ab$ are almost independent of~$T$, whereas that for $H \parallel c$ is weakly $T$-dependent, consistent with the $\chi(T)$ data in Fig.~\ref{Fig_M-T}(c).

\begin{figure}
\includegraphics[width=3.3in]{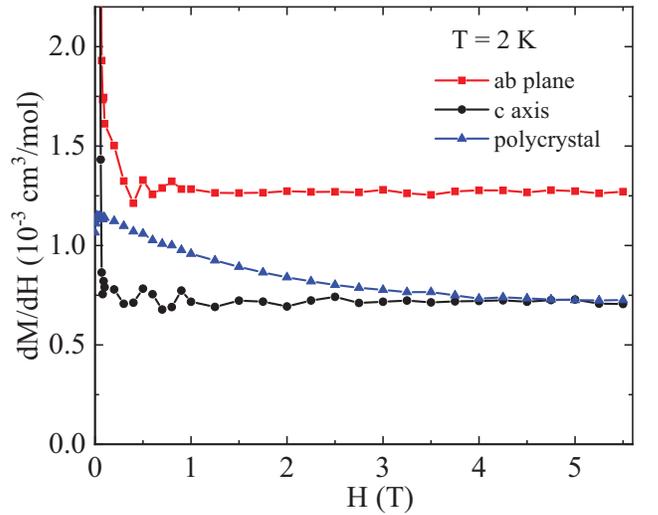}
\caption{Field derivative $dM/dH$ vs~$H$ obtained from the $M(H)$ data at $T=2$~K for SrCr$_2$As$_2$ in Fig.~\ref{Fig_Hysteresis_2K}.}
\label{Fig_dM-dH}
\end{figure}

According to the first-principles calculations in Ref.~\cite{Zhou2019}, the magnetocrystalline anisotropy energy (MAE) favors $c$-axis moment alignment by an amount
 \begin{eqnarray}
{\rm MAE([001]-[100]) = -0.20~meV/Cr~atom,}
  \end{eqnarray} 
consistent with the $c$-axis AFM ordering.
Taking the observed ordered moment \mbox{$\mu=1.9\,\mu_{\rm B}$/Cr~atom} from Ref.~\cite{Das2017}, in a \mbox{$c$-axis} magnetic field the spin-flop transition field $H_{\rm SF}$ at $T=0$ is calculated to be
 \begin{eqnarray}
H_{\rm SF}  = \frac{\rm MAE}{\mu} = 1.8~{\rm T}.  
  \end{eqnarray} 
The $M(H,T=2~{\rm K})$ data for $H\parallel c$ in Fig.~\ref{Fig_M-H}(b) show no evidence for a spin-flop transition at this field.  To look for a small anomaly at this field, Fig.~\ref{Fig_dM-dH} shows a plot of $dM(H)/dH$ versus $H$ at $T=2$~K and no evidence of a transition at 1.8~T is seen.  
\ This indicates that  no spin-flop transition at least up to 5.5 T, suggesting that  the MAE is greater than $|$--0.61$|$ meV/Cr atom based on Eq.~(2).  
%It is unclear why a spin-flop transition was not observed in Fig. 5(b) at the $c$-axis field.  The reason seems likely related to the unknown reason that $\chi_c$ at $T << T_{\rm N}$ in Fig. 3(c) is about half of $\chi_{ab}$ at $T << T_{\rm N}$, where the former value would be expected to be zero for a local-moment antiferromagnet with the same magnetic structure.  
The difference between the observed and the calculated values is important for further theoretical  investigation.
%   the theoretical those expected from the local-moment scenario are important to be investigated theoretically.
As shown above,  the $M(H,T=2~{\rm K})$ data for $H\parallel c$ have a nonzero slope and the $\chi_{c}(T)$ data in Fig.~\ref{Fig_M-T}(c) are nonzero, contrary to the expectation of  $\chi_c(T\to0)=0$ for a collinear $c$-axis antiferromagnet. 
In addition, we note that if the Cr moments are considered to be point dipoles, the magnetic ordering direction from the magnetic-dipole interaction  would be in the $ab$ plane instead of along the $c$ axis \cite{Johnston2016}. 
% These results suggest  that the magnetism of ${\rm SrCr_2As_2}$ is itinerant.
These differences from expectation for a local-moment system likely originate from the itinerant nature of the magnetism in SrCr$_2$As$_2$.
%  as discussed further below. 

% The formal oxidation state of a Cr atom in ${\rm SrCr_2As_2}$ is 2+ (3$d^4$ electron configuration).  
%   In the tetrahedral coordination of Cr by As, the two $e_g$ $d$ orbitals are lower in energy than the three  $t_{2g}$ orbitals.  
 %   In a local-moment picture, the high-spin state therefore has spin $S$ = 2 and the low-spin state has $S$ = 0.  
 %    In the former case, the ordered moment at $T$ = 0 K is expected to be $\mu$  = $g S \mu_{\rm B}$  = 4 $\mu_{\rm B}$/Cr, where $g$ = 2 is the spectroscopic-splitting factor and $\mu_{\rm B}$  is the Bohr magneton. 
%  However, the value of $\mu$ measured by neutron diffraction at $T$  = 12 K is instead the much smaller value 1.9(1) $\mu_{\rm B}$/Cr, indicating that the Cr magnetism is itinerant \cite{Das2017}.  The DFT calculations in Ref.~\cite{Zhou2019} also indicate that the magnetic ordering is itinerant.

\subsection{Heat capacity measurements}

\begin{figure}
\includegraphics[width=3.3in]{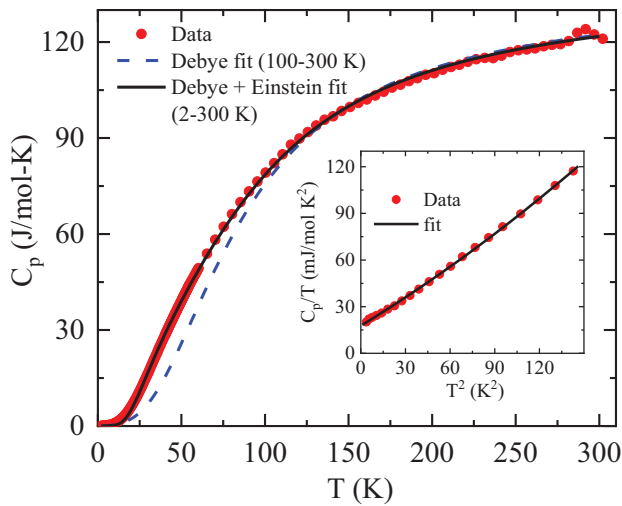}
\caption{Temperature $T$ dependence of the zero-field heat capacity $C_{\rm p}$ for a SrCr$_2$As$_2$ single crystal. The bump in the data at $\approx 290$~K is an experimental artifact. The data in the $T$ range 100--300~K are fitted by the Debye model using Eq.~(\ref{Eqs:Debye_Fit}) (dashed curve), whereas the data in the full $T$~range 1.8--300~K are better described by a sum of Debye and Einstein contribution using Eq.~(\ref{Eq:Debye and Einstein}) as shown by the solid curve in the figure. Inset: $C_{\rm p}/T$ vs $T^2$ at low temperatures along with a fit by Eq.~(\ref{Eq.CpFit1}) (solid curve). }
\label{Fig_Heat_Capacity}
\end{figure}

The zero-field heat capacity $C_{\rm p}(T)$ of single-crystal SrCr$_2$As$_2$ for $T=1.8$--300~K is plotted in Fig.~\ref{Fig_Heat_Capacity}. The $C_{\rm p}$ attains a value of $\approx 121$~J/mol\,K at $T = 300$~K, which is close to the classical Dulong-Petit high-$T$ limit $C_{\rm V} = 3nR = 124.71$ J/mol K associated with acoustic lattice-vibration modes, where $n = 5$ is the number of atoms per formula unit and $R$ is the molar-gas constant. The bump observed at $T \approx 290$~K in $C_{\rm p}(T)$ is due to melting of the Apiezon N-grease used for making thermal contact between the crystal and sample platform of the heat capacity puck and is hence an instrumental artifact.

The low-$T$ $C_{\rm p}(T)$ data for 1.8~K~$\leq T \leq$~12~K were analyzed using the relation
 \begin{eqnarray}
 C_{\rm p}(T) = \gamma T+ \beta T^3 +\delta T^5,
\label{Eq.CpFit1}
  \end{eqnarray} 
where $\gamma$ is the Sommerfeld electronic heat-capacity coefficient and $\beta$ and $\delta$ are coefficients associated with the low-$T$ lattice heat-capacity contribution. The inset of Fig.~\ref{Fig_Heat_Capacity} shows  $C_{\rm p}(T)/T$ versus $T^2$ along with the fit by Eq.~(\ref{Eq.CpFit1}). The fitted parameter values are \mbox{$\gamma = 17.7(3)$~mJ/mol\,K$^2$}, $\beta = 0.58(1)$~mJ/mol\,K$^{4}$, and $\delta = 0.8(1)~\mu$J/mol K$^6$. The value of $\beta$ was used to determine the Debye temperature $\Theta_{\rm D}$ according to
 \begin{eqnarray}
\Theta_{\rm D} = \bigg(\frac{12\pi^4Rn}{5\beta}\bigg)^{1/3} = 256(2)~{\rm K}.
\label{Eq.CpFit2}
  \end{eqnarray} 

The $C_{\rm p}(T)$ data for the full temperature range \mbox{1.8~K~$\leq T \leq$~300~K} were fitted by
 \begin{eqnarray}
C_{\rm p}(T) = \gamma T+ nC_{\rm V\,Debye}(T),
\label{Eqs:Debye_Fit}
  \end{eqnarray} 
where the Debye lattice heat capacity at constant volume is given by
 \begin{eqnarray}
C_{\rm V\,Debye}(T) = 9R \left(\frac{T}{\Theta_{\rm D}}\right)^3\int_{0}^{\Theta_{\rm D}/T}\frac{x^4e^x}{(e^x-1)^2} dx.
  \end{eqnarray} 
Here, we used $\gamma = 17.7$~mJ/mol\,K$^{2}$ obtained from the above low-$T$ data analysis and $C_{\rm V\,Debye}(T)$ was calculated using an accurate analytic Pad\'{e} approximant function~\cite{Goetsch_2012}.  As seen from Fig.~\ref{Fig_Heat_Capacity}, the $C_{\rm p}(T)$ data are only described by the Debye model in the temperature range  100--300~K with $\Theta_{\rm D} = 342(2)$~K\@. This value of $\Theta_{\rm D}$ is larger than the value of 256(2)~K determined above from the fit of the $C_{\rm p}(T)$ data at low $T$ by Eq.~(\ref{Eq.CpFit1}). 

A better fit to the $C_{\rm p}(T)$ data over the full temperature range was obtained as the sum of the electronic and the Debye and Einstein lattice contributions
 \begin{eqnarray}
C_{\rm p}(T) &=& \gamma T+ (1 - \alpha)C_{\rm V\,Debye} + \alpha C_{\rm V\,Einstein},
\label{Eq:Debye and Einstein}  
\end{eqnarray} 
where
 \begin{eqnarray}
C_{\rm V\,Einstein}(T/\Theta_{\rm E}) &=& 3R \left(\frac{\Theta_{\rm E}}{T}\right)^2\frac{e^{\Theta_{\rm E}/T}}{(e^{\Theta_{\rm E}/T} - 1)^2}
\label{Eq:Einstein}
\end{eqnarray} 
and $\Theta_{\rm E}$ as the Einstein temperature. The relative fraction of the Debye and Einstein phonon contributions is determined by the parameter $\alpha$\@. The best fit is obtained using the previous value \mbox{$\gamma = 17.7$~mJ/mol K$^{2}$,} together with $\Theta_{\rm D} = 434(3)$~K, $\Theta_{\rm E} = 119(1)$~K, and $\alpha = 0.37(1)$ as shown by the solid black curve in Fig.~\ref{Fig_Heat_Capacity}.
  A similar value \mbox{$\gamma = 19.3$~mJ/mol K$^{2}$} has been reported for BaCr$_2$As$_2$ \cite{Singh2009}.

Now we estimate the density of states  at the Fermi energy [${\cal D}(E_{\rm F})$] from the value of  \mbox{$\gamma = 17.7$~mJ/mol K$^{2}$}.
Using  the relation $\gamma$ = $\frac{1}{3}\pi^2k_{\rm B}^2{\cal D}(E_{\rm F})$, ${\cal D}(E_{\rm F})$ is calculated to be 7.53 states/eV f.u. for both spin directions.
  From a comparison with a bare band-structure value of 3.7 states/eV f.u. for both spin directions obtained from a first-principles calculation \cite{Zhou2019},   the effective mass of the carriers is suggested to be enhanced by a factor of $\sim$2 in the AFM state of SrCr$_2$As$_2$. 
A similar enhancement has been pointed out in BaCr$_2$As$_2$ \cite{Singh2009}.

Finally it is noted that, in the present analysis, we did not include the magnetic contribution to the heat capacity. It is known that in the absence of an excitation energy gap, AFM spin waves at low temperatures have a $T^3$ heat-capacity dependence if the spin waves are three-dimensional and a $T^2$ dependence if two-dimensional. 
   Since we have no information on the magnon dispersion relations, it was not possible to separate the contribution of spin waves from the contribution of the phonons to the $\beta$$T^3$ term in the low-$T$ heat capacity, so by necessity we ignored a possible contribution of 3D spin waves to this term. 
   However, in either 2D or 3D AFM spin waves, the presence would not affect the $\gamma$$T$ term in the heat capacity, which is the only term relevant to extracting the electronic contribution to the heat capacity.

\subsection{Electrical resistivity measurements}

% \red{To be added by Makariy Tanatar}

\begin{figure}
\includegraphics[width=\columnwidth]{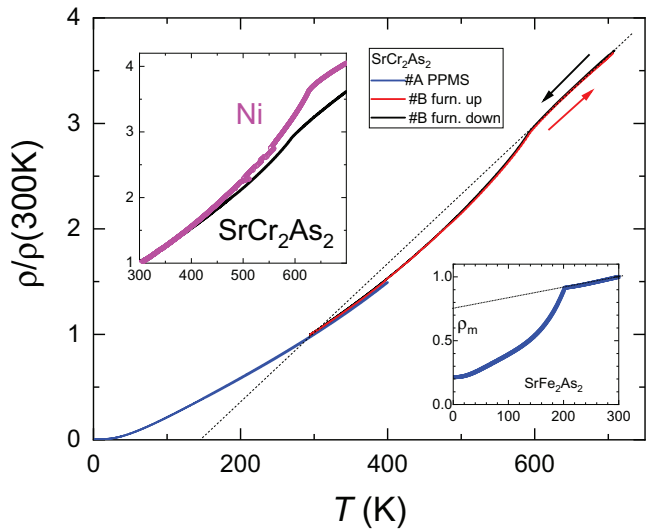}
\caption{ In-plane electrical resistivity $\rho_a$ of two samples of SrCr$_2$As$_2$ compiled from measurements in the PPMS (sample \#A, 1.8~K to 400~K) and in the furnace set-up (\#B, 300~K to 700~K). 
   Measurements on sample \#B were taken on warming (from 300~K to 700~K, red curve and red arrow) and on cooling from 700~K to 300~K (black curve and black arrow). 
    The resistivity was normalized by the value at room temperature, $\rho$(300~K). 
    This value was determined as $\rho_a$(300 K)~=~48~$\pm$~10~$\mu \Omega$$\cdot$cm. 
     The left inset compares high-temperature resistivity measurements of SrCr$_2$As$_2$ (black curve) with metallic Ni (magenta curve). 
    The magenta and black arrows indicate the temperatures of the ferromagnetic transition in Ni and the antiferromagnetic transition in SrCr$_2$As$_2$, respectively. 
     The right inset shows the $T$ dependence of  the normalized $\rho_a$/$\rho$(300~K) for SrFe$_2$As$_2$ from Ref. \cite{anisotropy2}.
    The dashed line shows a linear extrapolation of the normalized $\rho(T)$ curve from above the transition to $T=$0, showing the contribution of magnetic scattering in spin-disordered phase, similar to the main panel.   
}
\label{rho}
\end{figure}

   Temperature-dependent electrical resistivity measurements were performed on several samples with current along the $a$-direction in the plane, $\rho_a$. Measurements in the temperature range 1.8 to 400~K, performed on two different samples, were combined with measurements in the range 300~K to 700~K, performed on two other samples. 
  The curves for the samples were normalized by the value of the resistivity at room temperature, $\rho$(300~K). 
The combined data are presented in the main panel of Fig. 8.
   The data in two different set-ups match smoothly in the range of overlap, 300 to 400~K. 
   The $\rho_a (T)$ curve shows a metallic resistivity increase on warming in the whole temperature range up to 700~K. 
   The $\rho_a(T)$ is linear in the paramagnetic state above $T_{\rm N}$, and shows clear positive curvature below $T_{\rm N}$. 
   This resistivity behavior below $T_{\rm N}$ is typical for metals with contribution of spin-disorder scattering in the paramagnetic state \cite{wilding}. 

The data for SrCr$_2$As$_2$ are very similar to those in BaCr$_2$As$_2$ single crystals \cite{Filsinger2017, Richard2017} and EuCr$_2$As$_2$ \cite{Eu}, though measurements in these compounds were not made to high enough temperatures to reach $T_{\rm N}$. They also bear similarity to $\rho_a(T)$/$\rho$(300 K) for SrFe$_2$As$_2$ \cite{anisotropy2} (right inset of Fig.~\ref{rho}).  
In the case of SrFe$_2$As$_2$, and as usual for magnetic scattering \cite{MS}, the linear extrapolation of $\rho(T)$ curve above $T_{\rm N}$ to $T$ = 0 gives a big positive offset on the resistivity axis. 
A similar extrapolation in SrCr$_2$As$_2$ (dashed line in the main panel of Fig.~\ref{rho}) gives a negative value, indicating a significantly stronger interaction of the charge carriers with the magnetic subsystem in the iron compounds.

   Interestingly, the resistivity curve does not show any noticeable increase below $T_{\rm N}$ due to the appearance of magnetic superzone boundaries. This type of behavior is more common for ferromagnets. 
   In the left inset of Fig.~\ref{rho},  we compare our measured electrical resistivity of ferromagnetic Ni with that of SrCr$_2$As$_2$. 
    Both samples show a clear downward deviation from the resistivity in the paramagnetic state. 
   The derivative of the resistivity defines the ferromagnetic Curie temperature $T_{\rm C}$ of Ni as 630~K, in good agreement with a suggested value of 631~K \cite{Ni}. 
   In Ni there is no superzone boundary formation due to $Q=0$ modulation in the ferromagnetic state. 
   Superzone boundaries are formed at the points of electron energy bands crossing new magnetic Brillouin zone boundaries in the magnetically-ordered state \cite{wilding}. 
   Modulation with a $Q=0$ wavevector does not lead to a new periodicity in a ferromagnet, and as a result this effect is absent here. 
 To understand the role of the magnetic superzone boundaries in SrCr$_2$As$_2$, 
% Unexpectedly, there is no clear resistivity increase below $T_{\rm N} \approx$ 600~K in SrCr$_2$As$_2$. To understand the role of the magnetic superzone boundaries in SrCr$_2$As$_2$
 we refer to the band structure calculations \cite{Zhou2019,Hamri2021},  suggesting that the Fermi surface has the shape of a warped cylinder. 
   This shape of the Fermi surface is indeed found experimentally in an ARPES study of the closely related BaCr$_2$As$_2$ \cite{Nayak2017}. 
   The three-dimensional character of G-type magnetic ordering makes a poor match with the Fermi surface of this kind, with the magnetic Brillouin zone affecting very small areas of the Fermi surface. 
%     Taking the magnetic susceptibility decrease below $T_{\rm N}$ as a proxy of the decrease in the  density of states [Fig. 3(a)] (this estimate assumes a paramagnetic character of the spin susceptibility \red{even though it contains other contributions}),  we arrive at a decrease on the order of 25\%. 
Compared with the small change of the density of states below $T_{\rm N}$, clearly the loss of spin-disorder scattering provides a much bigger effect.
Considering that the contribution of spin-disorder scattering vanishes with magnetic ordering, the curvature of the resistivity curve below $T_{\rm N}$ reflects mainly a build-up of the magnetic order parameter [see Fig.~\ref{fig:NMR}(d) below] and residual magnetic entropy \cite{Rh}. 

\begin{figure}
\includegraphics[width=\columnwidth]{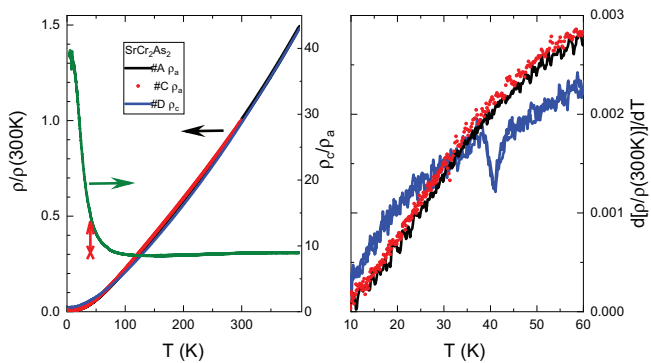}
\caption{The left panel compares the in-plane $\rho_a$ (black and red curves for samples \#A and \#C), and interplane $\rho_c$ (blue curve for sample \#D) resistivity of SrCr$_2$As$_2$. 
The black curve is slightly difficult to see due to the overlap with other two curves. 
      The temperature dependence of the resistivity anisotropy $\rho_c/\rho_a$ is also plotted by the green curve in the left panel.  
 For a broad temperature range the anisotropy remains rather constant, but increases notably below 50~K. 
 The red cross-arrow indicates the position of the anomaly in temperature-dependent resistivity derivative of $\rho_c(T)$ shown in the light panel where the temperature dependence of $d$[$\rho$/$\rho$(300~K)]/$dT$ is plotted. 
}
\label{caxis}
\end{figure}

In Fig.~\ref{caxis}, we compare the in-plane $\rho_a$ and interplane $\rho_c$ resistivity of SrCr$_2$As$_2$. 
The $\rho_c$ data were taken in two-probe mode. 
The measurements were made from 1.8~K to 400~K. 
The inter-plane resistivity at room temperature was determined as $\rho_c$(300 K)~=~430 $\pm$ 100~$\mu \Omega$$\cdot$cm.
% The resistivity anisotropy $\rho_c$/$\rho_a$ is about 9 $\pm$ 2 at room temperature, but increases notably below 50~K. 
%The rather low anisotropy is determined by strong warping of the Fermi surface along $c$ axis and suggests that the inter-plane transport is coherent, as indeed found in experiment (the left panel of Fig. 9).}
%The resistivity anisotropy is small at room temperature, about 9$\pm$2, consistent with strong warping of the Fermi surface, but increases notably below 50~K. 
%    A slight anomaly can be found in the $\rho_c(T)$ temperature derivative (right panel in Fig.~\ref{caxis}), the origin of which remains unknown.
%  No anomalies are observed in this temperature range in other measurements. 
The resistivity anisotropy $\rho_c/\rho_a$ is about 9  $\pm$ 2 at room temperature, and its temperature dependence is shown by the green curve in the left panel of Fig. 9 (the right axis). 
   The anisotropy remains nearly constant down to 50~K, and then shows an about four-fold increase on further cooling. Interestingly, the onset of anisotropy increase correlates with a slight anomaly in the temperature dependence of the inter-plane resistivity derivative (see the right panel in Fig. 9). 
    The origin of this anomaly remains unknown and it is not observed in other measurements. 
   % The rapid increase of the anisotropy is suggestive of some magnetic structure change. 

   \subsection{NMR measurements}
     \subsubsection{ $^{75}$As NMR spectra of SrCr$_2$As$_2$ single crystal}

% \begin{figure}[tb]
% \includegraphics[width=\columnwidth]{Fig1-1} 
% \caption{(a) Crystal and magnetic structures of SrCr$_2$As$_2$  }
% \label{fig:NMR}
% \end{figure}   

 \begin{figure*}[tb]
\includegraphics[width=18cm]{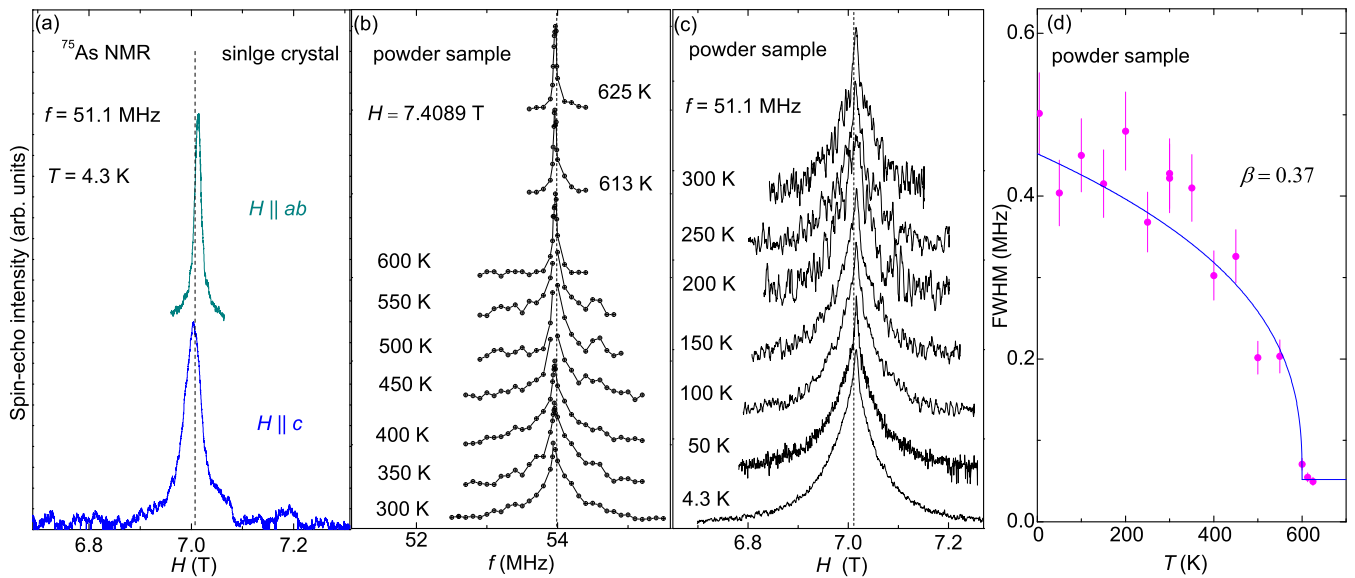} 
\caption{ (a) Field-swept $^{75}$As-NMR spectra of a SrCr$_2$As$_2$ single crystal measured at $f$ = 51.1 MHz and $T$ = 4.3 K, for $H$ $\parallel$ $c$ axis and $H$ $\parallel$ $ab$ plane.
  The vertical dashed line represents the zero-shift position.
(b)~Temperature dependence of the $f$-swept $^{75}$As-NMR spectrum for SrCr$_2$As$_2$ powder sample measured at $H$ = 7.4089 T above $T$ = 300 K.
The vertical dashed line represents the zero-shift position.
(c)~Temperature dependence of the $H$-swept $^{75}$As-NMR spectrum for a SrCr$_2$As$_2$ powder sample measured at $f$ = 51.1 MHz below $T$ = 300 K.
The vertical dashed line represents the zero-shift position.
(d)~Temperature dependence of the FWHM of the $^{75}$As-NMR spectrum for the SrCr$_2$As$_2$ powder sample.
     The solid line is calculated with  FWHM($T$) = FWHM(0)(1 $-$ $T/T_{\rm N}$)$^{\beta}$ + $C$ where  $\beta$ = 0.37,  $T_{\rm N}$ = 600 K, FWHM(0)~=~0.4~MHz and  $C$~=~0.052~MHz.
}
\label{fig:NMR}
\end{figure*}   

   Figure~\ref{fig:NMR}(a) shows the field-swept $^{75}$As-NMR spectra of a SrCr$_2$As$_2$ single crystal measured at frequency $f$~=~51.1~MHz in the AFM state ($T$ = 4.3 K) for $H$ $\parallel$ $c$ axis and $H$ $\parallel$ $ab$ plane.
   For the $I$ = 3/2 nuclei, one may expect, in the presence of quadrupole effects, three spectral lines: one central ($I_z$ = 1/2 $\leftrightarrow$ --1/2) and two satellite transitions ($I_z$ = 3/2 $\leftrightarrow$ 1/2 and --3/2 $\leftrightarrow$ --1/2). 
    The observed $^{75}$As-NMR spectra for both $H$~$\parallel$~$c$ axis and $H$~$\parallel$~$ab$ plane only show a single line around the Larmor field [dashed line in Fig.\ \ref{fig:NMR}(a)].
   The observation of the NMR line around the Larmor field indicates that there is no large internal magnetic field at the As site produced by the Cr ordered moments in the AFM state.
   The absence of clear satellite lines suggests that the quadrupolar interaction is quite small at the As site and we estimate the upper limit of the quadrupole interaction frequency $\nu_{\rm Q}$ to be $\approx$ 0.1~MHz. 
     The full width at half maximum  (FWHM)  of the line for $H$ $\parallel$ $c$ is  $\approx$ 0.41~kOe which is greater than 0.16~kOe for $H$ $\parallel$ $ab$.
      No change in the peak positions and FWHMs were observed up to 100~K. 
     Above 100 K, the NMR measurements on the single crystal became difficult due to weak signal intensity.

   The internal field at the $^{75}$As site can be analyzed by taking the crystal symmetry into consideration, which has been adopted in an analysis of the hyperfine field at the $^{75}$As site in BaFe$_2$As$_2$ by Kitagawa et al. \cite{Kitagawa2008}.  
    According to their analysis, for a G-type AFM spin structure, the internal field at the As site is zero due to a perfect cancellation of the off-diagonal hyperfine field produced by four in-plane nearest-neighbor Cr spins when the spin moments are parallel to the $c$ axis. 
    Thus, the spin components along this axis do not produce any shift  in the $^{75}$As NMR spectra. 
    On the other hand, if there were $ab$ plane components of the ordered Cr spin, they would  produce an internal field perpendicular to the $c$ axis at the $^{75}$As site \cite{Kitagawa2008}. 
    Thus one may expect the peak position to be shifted due to the internal field. 
    However, according to the ND results,  the AFM state is G-type where the  Cr$^{2+}$ ordered moments align along the $c$ axis. 
    Therefore, one expects a zero internal field at the As site in the AFM state of SrCr$_2$As$_2$ and should observe the NMR line around the Larmor field.
    This is consistent with what we observed in the $^{75}$As NMR spectrum measurements, confirming  the G-type AFM state in SrCr$_2$As$_2$.

 \subsubsection{$^{75}$As NMR spectra of SrCr$_2$As$_2$ powder sample}

     To overcome the difficulties for the NMR measurements at higher temperatures above 100 K with the SrCr$_2$As$_2$ single crystal, we used a polycrystalline  sample which makes NMR measurements possible up to 625~K, which is above the N\'eel temperature.
   Figures \ \ref{fig:NMR}(b) and \ \ref{fig:NMR}(c) show the temperature dependence of the $f$-swept $^{75}$As NMR spectrum of polycrystalline sample measured at $H$ = 7.4089 T for $T \geq$ 300~K, and the temperature dependence of the $H$-swept $^{75}$As NMR spectrum measured at $f$ = 51.1 MHz for $T \leq$ 300~K, respectively.
  As in the case of the single crystal, a single peak expected for a small quadrupolar interaction was observed in the whole temperature range from 625~K down to 4.3~K.
  Although no clear change in the peak position of the NMR spectra is observed, as the temperature decreases, the NMR spectra broaden  below $\approx$ 600~K, which evidences the magnetic ordered state below this temperature.

    Figure~\ref{fig:NMR}(d) shows the temperature dependence of the FWHM of the NMR spectra where the FWHM increases gradually with decreasing temperature.
   Although the internal field at the As site is expected to be zero in the AFM state  as described above,  a distribution of the Cr ordered moments and/or distribution of the transferred hyperfine coupling constant would broaden the NMR spectra \cite{Ding2017}.
     Since the temperature dependence of the FWHM for the NMR line in the case of polycrystalline samples in a magnetically-ordered state reflects the temperature dependence of the order parameter (the magnitude of the Cr$^{2+}$ ordered moments), the $T$ dependence of the FWHM below 600 K indicates  a second-order phase transition.
    In general, one can estimate a critical exponent for the AFM transition from the temperature dependence of the FWHM. 
    However, our experimental data are somewhat scattered,  so we do not estimate the critical exponent from our experimental data.
   Instead, we calculated the temperature dependence of the FWHM  based on the ND measurements \cite{Das2017}.
   The solid curve in Fig.~\ \ref{fig:NMR}(d) is the calculated result of FWHM($T$) = FWHM(0)(1 $-$ $T/T_{\rm N}$)$^{\beta}$ + $C$ with $\beta$~=~0.37 reported from the ND measurements. 
  Here we used  FWHM(0) = 0.4~MHz and  $C$~=~0.052~MHz which is the FWHM just above $T_{\rm N}$.
   As shown, the curve with $\beta$~=~0.37 reasonably reproduces the experimental data.
   Since the value of $\beta$ is close to  0.33--0.367 for  three-dimensional (3D) Heisenberg, 0.31--0.345 for 3D XY, and 0.3--0.326 for 3D Ising models  but much greater than 0.125 for the two-dimensional Ising model \cite{Nath2009}, the results  suggest a three-dimensional nature of the magnetism in SrCr$_2$As$_2$, as found from the ND measurements \cite{Das2017}.

\subsubsection{$^{75}$As spin-lattice relaxation rate 1/$T_1$}
    
\begin{figure}[tb]
\includegraphics[width=\columnwidth]{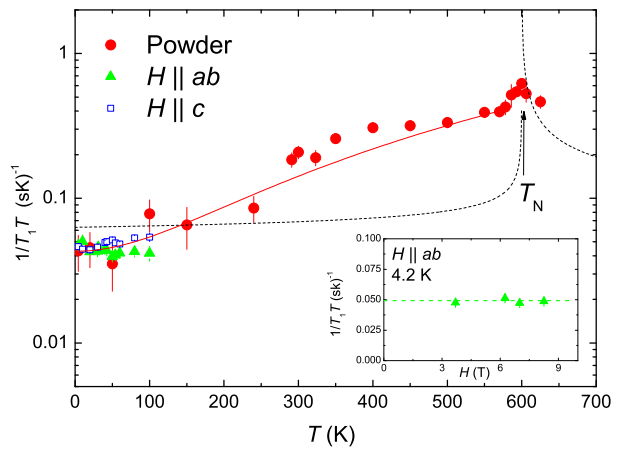} 
\caption{Temperature dependence of the nuclear spin-lattice relaxation rate 1/$T_1$ divided by $T$ for both the SrCr$_2$As$_2$ single crystal and powder sample. The black dashed and red lines are the calculated results based on the SCR theory and the two-magnon scattering model, respectively (see text). 
 Inset: Magnetic field  dependence of 1/$T_1T$ at $T$ = 4.2 K for $H$~$||$~$ab$. The dashed  horizontal line represents the average value.
}
\label{fig:T1}
\end{figure}

     The AFM phase transition has also been detected from the $^{75}$As spin-lattice relaxation rate (1/$T_1$) measurements.
    Figure\ \ref{fig:T1} shows the temperature dependence of 1/$T_1T$ of the polycrystalline SrCr$_2$As$_2$ sample for $T$ = $4.2 - 625$ K, together with the  1/$T_1T$ results for the SrCr$_2$As$_2$ single crystal at low temperatures below 100~K for $H$ $\parallel$ $c$ axis and $H$ $\parallel$ $ab$ plane.
    1/$T_1T$ increases as temperature decreases from 625 K, and shows a peak at $T_{\rm N}$ $\approx$ 600 K, which is due to the critical slowing down of spin fluctuations for a second-order phase transition.
%     To see the magnetic phase transition clearly, we show the temperature dependence of 1/$^{75}T_1T$ around $T_{\rm N}$ in the inset of Fig.\ \ref{fig:T1}.
    Below $T_{\rm N}$, with decreasing temperature, 1/$T_1T$ decreases gradually from 550~K to  350~K, then shows a relatively steep decrease below 300~K and finally exhibits a 1/$T_1T$ = constant behavior below 100 K.

   In the case of AFM metals, 1/$T_1T$ is given by 
\begin{eqnarray}
1/T_1T = (1/T_1T)_{\rm const} + (1/T_1T)_{\rm AFM} 
\label{eqn:T1}
\end{eqnarray}
where $(1/T_1T)_{\rm const}$ is the temperature-independent value originating from  conduction carriers and the second term is due to AFM fluctuations.
    In the framework of weak itinerant antiferromagnets, the self-consistent renomalization (SCR) theory predicts the following relations \cite{SCR1}
\begin{eqnarray}
(1/T_1T)_{\rm AFM} = \frac{a}{\sqrt{(T-T_{\rm N})^{1/2}}} ~~ (T > T_{\rm N}) \\
% = \frac{b}{\sqrt{(T-T_{\rm N})^{\beta}}} ~~ (T < T_{\rm N}) \\
= \frac{b}{M_{\rm Q}(T)} ~~ (T < T_{\rm N}) .
\label{eqn:T1_2}
\end{eqnarray}
Here $M_{\rm Q}(T)$ represents the temperature dependence of the staggered moments in the antiferromagnetic state.
Although we tried to fit the data by changing the parameters, we were not able to reproduce the experimental data with the model. 
The typical results are shown by the black dashed lines in Fig.~11  calculated with the parameters $T_{\rm N}$ = 600 K, $a$ = 1.5 (sK)$^{-1}$ , $b$ = 2$\times$10$^{-2}$ (sK)$^{-1}$ and  $ (1/T_1T)_{\rm const}$ = 0.043 (sK)$^{-1}$. 
Here we used  the calculated FWHM$(T)$ as $M_{\rm Q}(T)$.
% Utilizing possible values of $T_{\rm N}$ = 600 K, $a$ = 1.5 (sK)$^{-1}$ , $b$ = 6$\times$10$^{-5}$ (sK)$^{-1}$ and  $ (1/T_1T)_{\rm const}$ = 0.043 (sK)$^{-1}$, and also with an assumption of the temperature dependence of FWHM for  $M_{\rm Q}(T)$,  . 
% As shown, the experimental data are not reproduced by the theory. 
   Then,  we consider another possibility for $(1/T_1T)_{\rm AFM}$. 
   When 1/$T_1T$ is mainly driven by scattering of magnons, often observed in antiferromagnetic insulators, $(1/T_1T)_{\rm AFM}$ is expected to follow a $T^2$ power-law temperature dependence due to a two-magnon Raman process \cite{Beeman1968}. 
    The red solid line in Fig.~11 is the result calculated  with $1/T_1T$ =  0.043  + 1.2$\times$10$^{-6}$$T^2$ (sK)$^{-1}$, which reproduces the  experimental data better than the fit with the SCR theory.
%  in the temperature range of $T$ $\sim$ $150 - 550$ K.
A similar analysis of the temperature dependence of 1/$T_1T$ has been reported in the antiferromagnetic metal K-doped BaMn$_2$As$_2$ \cite{Yeninas2013}. 
    These results suggest that the nuclear relaxation process in the AFM state in the temperature region $T$ $\sim$ $100 - 550$ K mainly originates from magnon scattering and also suggest a relatively strong localized nature of the Cr ordered moments in the AFM state of SrCr$_2$As$_2$. 
    On the other hand, the $T$-independent 1/$T_1T$ = 0.043 (sK)$^{-1}$ observed below 100 K indicates that the nuclear relaxation in the temperature range is dominated by the relaxation process due to the conduction carriers, evidencing the metallic state from a microscopic point of view. 
  The duality of localized and itinerant natures may originate from $d$ electrons on the different 3$d$ orbitals of the Cr ions in  SrCr$_2$As$_2$.

   Since the $T$-independent 1/$T_1T$ observed at low temperatures  is in general proportional to the square of  the density of states  at the Fermi level ${\cal D}(E_{\rm F})$ as \cite{Narath_Hyper} $(T_{1}T)^{-1}=4\pi\gamma_{\rm N}^2\hbar k_{\rm B}A_{\rm hf}^2 {\cal D}^2(E_{\rm F})$, where $k_{\rm B}$  is Boltzmann's constant and $A_{\rm hf}$ is a hyperfine coupling constant, these results indicate that ${\cal D}(E_{\rm F})$ is independent of temperature at least below 100 K.
    This is in contrast to the recent Raman spectroscopy measurements \cite{Kaneko2021} which suggest a continuous decrease of the density of states at $T$ $\textless$ 250 K.
     Similar $T$-independent 1/$T_1T$ behaviors with 1/$T_1T$ $\sim$ 0.043 (sK)$^{-1}$ are also observed in the SrCr$_2$As$_2$ single crystal below $\sim$ 100 K for two different magnetic field directions  $H$~$\parallel$~$c$ axis and $H$~$\parallel$~$ab$, as shown in Fig.~\ref{fig:T1}.  
     These results again indicate the $T$-independent behavior of the ${\cal D}(E_{\rm F})$ below 100 K.
In addition,  the similar values of 1/$T_1T$ for $H$ $\parallel$ $c$  and $H$ $\parallel$ $ab$  suggest no obvious anisotropy in the magnetic fluctuations due to conduction carriers in this material.

 It is noted that we did not attribute the decrease in 1/$T_1T$ from $T_{\rm N}$ to 100 K to the reduction of  ${\cal D}(E_{\rm F})$.
  If the reduction of 1/$T_1T$ from 0.2 (sK)$^{-1}$ at 300 K to 0.043 (sK)$^{-1}$ at 4 K is due to the reduction of ${\cal D}(E_{\rm F})$, the ${\cal D}(E_{\rm F})$ must be reduced by a factor of $\approx$ 2.2 at low temperatures since 1/$T_1T$ is proportional to ${\cal D}^2(E_{\rm F})$.  
This seems to be inconsistent with  the observed $T$ dependence of $\chi_c$ and $\chi_{ab}$.
Assuming a Wilson ratio of 1, the Pauli paramagnetic susceptibility is estimated to be 2.43$\times$10$^{-4}$ cm$^3$/mol from the Sommerfeld electronic heat-capacity coefficient $\gamma$  = 17.7 mJ/mol K$^2$ (see Sec.~III~C). 
   This corresponds to $\approx$ 32 \% and $\approx$ 18 \% of the observed $\chi_c$ and $\chi_{ab}$, respectively below 300 K. 
   Therefore, although we do not know the origin of $\chi_c$ and $\chi_{ab}$ below $T_{\rm N}$, one may expect to observe the change in  $\chi_c$ and $\chi_{ab}$ if ${\cal D}(E_{\rm F})$ decreases with decreasing $T$.   
   However, as shown in Fig.~3(c),  the $\chi_c$ and $\chi_{ab}$ do not show an obvious reduction with decreasing $T$. 
    These results indicate no obvious change in ${\cal D}(E_{\rm F})$ with decreasing $T$. 

 It is also important to point out that the Raman spectroscopy measurements suggested the reduction of ${\cal D}(E_{\rm F})$  with increasing magnetic field \cite{Kaneko2021}. 
    To check this possibility, we measured the $H$ dependence of 1/$T_1T$ at 4.2 K for $H$ $||$ $ab$. 
   As shown in the inset of Fig.~\ref{fig:T1}, we found no obvious change in 1/$T_1T$ from 3.7 T to 8.25 T, indicating that the ${\cal D}(E_{\rm F})$ is nearly independent of $H$ up to 8.25 T.

\subsubsection{ $^{53}$Cr zero-field NMR in the antiferromagnetic state}

 \begin{figure}[tb]
\includegraphics[width=\columnwidth]{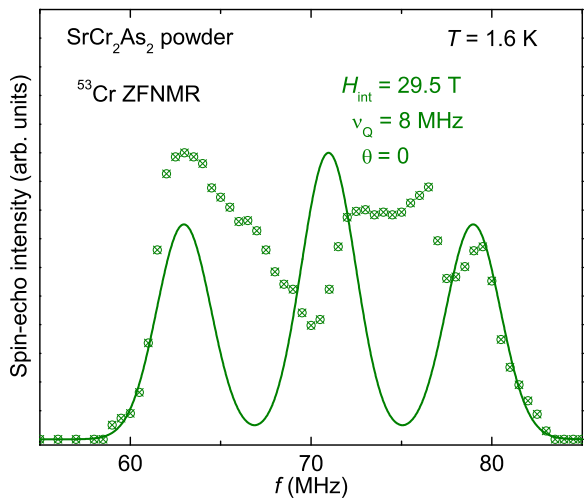} 
\caption{Zero-field $^{53}$Cr NMR spectrum at $T$ = 1.6 K.
The green curve is the calculated $^{53}$Cr-NMR spectrum using the parameters $B_{\rm int}^{\rm Cr}$ = 29.5 T,  $\nu_{\rm Q}$ = 8 MHz, and $\theta$ = 0.
}
\label{fig:Cr}
\end{figure}    
    
    We tried to observe zero-field Cr-NMR signals in the single crystals, but no signals were detected.  
However,  we succeeded in observing the $^{53}$Cr-NMR signal under zero magnetic field in  the SrCr$_2$As$_2$ powder sample at $T$~=~1.6~K in the  frequency range of $f$ $\sim$ 59--82~MHz as shown in Fig.~\ref{fig:Cr}, although the signal intensity was very small due to a low natural 9.5\% abundance of $^{53}$Cr and the broad spectrum.
%,  we succeeded in observing the $^{53}$Cr-NMR signal under zero magnetic field in SrCr$_2$As$_2$ at $T$ = 1.6~K in a frequency range of $f$ $\sim$ $59 - 82$~MHz as shown in Fig.\ \ref{fig:Cr}.
     This is direct evidence of the magnetic ordering of Cr moments in the AFM state.
 %     The zero-field $^{53}$Cr NMR spectrum measured at $T$ = 1.6 K is shown in Fig.\ \ref{fig:Cr}. 
     Since the signal intensity was very small even at the lowest temperature of 1.6 K, it is hard  to measure the spectrum precisely, making analysis of the spectrum difficult.
    Nevertheless, we tried to fit the spectrum.  
    The green curve in Fig.~\ref{fig:Cr} is the calculated $^{53}$Cr-NMR spectrum with a set of parameters of the internal magnetic field at the Cr site $|H_{\rm int}^{\rm Cr}|$ = 29.5~T, Cr quadrupolar interaction frequency  $\nu_{\rm Q}$~=~8~MHz, and $\theta$ = 0.
    Here $\theta$ represents the angle between  $H_{\rm int}^{\rm Cr}$ and the principle axis of the EFG tensor at the Cr sites.  
    Since the Cr site has a local fourfold symmetry around the $c$ axis in the tetragonal ThCr$_2$Si$_2$-type crystal structure, the principal axis of the EFG at the Cr site is parallel to the $c$ axis. 
    Therefore, since the Cr ordered moments align along the $c$ axis under zero magnetic field from the ND measurements  \cite{Das2017}, we used $\theta$ = 0 for the simulation.
      As shown in Fig.\ \ref{fig:Cr}, the observed spectrum was not reproduced well by the simulated spectrum, so we are not able to discuss  much about the parameters obtained from the simulation.  

   However, we consider that the frequency range we observed $^{53}$Cr-NMR signals provides some information about the magnitude of the Cr ordered moments.
    In general, the $^{53}$Cr-NMR  resonance frequency in magnetic materials is closely related to  the total spin value of the 3$d$ electrons on Cr ions since the resonance frequency is proportional to $H_{\rm int}^{\rm Cr}$ which is proportional to the Cr ordered moments $\langle \mu \rangle$ via the hyperfine interaction.
   Therefore, one may observe $^{53}$Cr NMR  in different frequency ranges with different ionic states of Cr ions. 
   For example, $^{53}$Cr-NMR signals for Cr$^{4+}$ with $S$ = 1 (3$d^2$) in the FM half-metal CrO$_2$ are observed around $f$ = 26--37~MHz \cite{Nishihara1972,Shim2007, Takeda2013},  while the signals are located around  65--70~MHz for Cr$^{3+}$ with $S$ = 3/2 (3$d^3$)  in insulating antiferromagnets Cr$_2$O$_3$ \cite{Rubinstein1964,Takeda2013}, YCrO$_3$ \cite{Jedryka1984,Takeda2013} and CuCrO$_2$ \cite{Smolnikov2015}.  
    $^{53}$Cr-NMR signals were also observed around 44--57~MHz for  $\langle \mu \rangle$ = 0.76--0.97 $\mu_{\rm B}$ in Cr-based molecular rings \cite{Micotti2006}.  
   Therefore, the resonance frequency $f$ = 59--80  MHz  observed in SrCr$_2$As$_2$ suggests that the Cr ordered moments have spin $\lesssim$ $S$ = 3/2 ($\mu$ = 3 $\mu_{\rm B}$).
    This is  less than the value of 4 $\mu_{\rm B}$ expected for Cr$^{2+}$ ($S$~=~2) ions in localized systems, which is consistent with the itinerant nature of the antiferromagnetism in SrCr$_2$As$_2$.          
   The hyperfine coupling constants for Cr ions have been reported to be $A_{\rm hf}$  = $-$11.7 to $-$12.7 T/$\mu_{\rm B}$ in Cr-based magnetic compounds \cite{Jedryka1984,Garlatti2020}. 
   Using   $H_{\rm int}^{\rm Cr}$ = 29.5 T and  $<$$\mu$$>$ = 1.9~$\mu_{\rm B}$/Cr and 2.26--2.4 $\mu_{\rm B}$/Cr from the ND measurements \cite{Das2017} and the first principle density-functional calculations \cite{Zhou2019,Hamri2021}, respectively, $|$$A_{\rm hf}$$|$ =  $H_{\rm int}^{\rm Cr}$/$<$$\mu$$>$  is estimated to be  15.5~T/$\mu_{\rm B}$ and 12.3 to 13.1~T/$\mu_{\rm B}$, respectively. 
    Although $|$$A_{\rm hf}$$|$  = 15.5 T/$\mu_{\rm B}$ is slightly greater than the reported value,  $|$$A_{\rm hf}$$|$  = 12.3 to 13.1 T/$\mu_{\rm B}$ is reasonably close to it.  These results indicate reduced ordered moments of Cr ions compared to the above local-moment value of 4 $\mu_{\rm B}$/Cr in SrCr$_2$As$_2$. 

% are consistent with the reduced ordered moments of Cr ions from the above local-moment value of 4 $\mu_{\rm B}$/Cr in SrCr$_2$As$_2$. }

%     Such reductions of the Cr ordered moments were reported from the ND measurements (1.9(1) $\mu_{\rm B}$ at $T$ = 12 K \cite{Das2017}) and also from the first-principles density-functional calculations (2.4 $\mu_{\rm B}$ \cite{Zhou2019} and 2.26 $\mu_{\rm B}$ \cite{Hamri2021}). 

\section{Summary} 
      The electronic and magnetic properties of itinerant antiferromagnetic SrCr$_2$As$_2$ with single crystalline and polycrystalline forms have been studied by a variety of measurements including electrical resistivity $\rho$, heat capacity $C_{\rm p}$, magnetic susceptibility $\chi$ versus temperature~$T$ and magnetization $M$ versus applied magnetic field $H$ isotherm measurements from a macroscopic point of view as well as  $^{75}$As and $^{53}$Cr NMR measurements from a microscopic point of view  in a wide temperature range of $T$ = 1.6--900 K. 

    The metallic ground state was directly evidenced by the electrical-resistivity and heat-capacity measurements as well as NMR measurements. 
   From the value of the Sommerfeld coefficient of the electronic heat capacity  \mbox{$\gamma = 17.7$~mJ/mol K$^2$},  the density of states at Fermi energy
  ${\cal D}(E_{\rm F})$ in the AFM state is estimated to be  7.53~states/eV~f.u.~for both spin directions, which is almost twice the bare ${\cal D}(E_{\rm F})$ estimated from first-principles calculations, suggesting an enhancement of the conduction carrier mass by a factor of two in the AFM state. 
   In addition, the temperature and magnetic-field dependences of the $^{75}$As spin-lattice relaxation rate divided by $T$, 1/$T_1T$,  at low temperatures indicate that the  ${\cal D}(E_{\rm F})$ is nearly constant at least below 100 K and is independent of $H$ up to 8.25 T, in contrast to the recent report from  the recent Raman spectroscopy measurements \cite{Kaneko2021} which suggest a continuous decrease of the ${\cal D}(E_{\rm F})$ below $T$ $\textless$ 250 K and also with increasing $H$.        
 
    The $\rho(T)$ is found to show $T$-linear behavior above $T_{\rm N}$ and exhibits  positive curvature below $T_{\rm N}$ where significant loss of spin-disorder scattering upon magnetic ordering is observed. 
   The $T$ dependence of $\rho(T)$ is found to  bear some similarity to the parent compounds of iron-based superconductors BaFe$_2$As$_2$ and SrFe$_2$As$_2$, but not CaFe$_2$As$_2$ \cite{anisotropy2}. 
   The resistivity anisotropy of the compound remains moderate $\rho_c/\rho_a \sim$9 through most of the magnetically ordered phase but shows rapid increase below 50~K.
% , the origin of which remains to be determined. 

     The AFM ordering temperature $T_{\rm N}$ = 615(15) K has been detected from  a clear peak  in $d(\chi T)/dT$, a slope discontinuity in $\rho(T)$, a sudden change in $^{75}$As NMR line-width and also a peak in the $T$ dependence of 1/$T_ 1$. 
      The observation of the $^{75}$As NMR spectrum around the zero-shift position (Larmor field) for both single-crystal and powder samples  in the AFM state below $T_{\rm N}$ indicates  a G-type AFM spin structure. 
    From the $\chi(T)$ measurements on  SrCr$_2$As$_2$ single crystal under the two different magnetic field directions $H$ $||$ $ab$ and  $H$ $||$ $c$ in the AFM state,  the Cr ordered moments are shown to align along the $c$ axis in the G-type AFM state, consistent with the results from the previous neutron-diffraction measurements.   
   The temperature dependence of 1/$T_1$ in the AFM state cannot be reproduced by the SCR theory for weak itinerant antiferromagnets, but it was found that a magnon-scattering model often observed in  antiferromagnetic insulators roughly reproduces the experimental data.
  The result suggests the localized nature of the Cr ordered moments in the metallic AFM state of SrCr$_2$As$_2$, showing the duality of  localized and itinerant natures which could originate from  $d$ electrons on the different 3$d$ orbitals of the Cr ions. 
   Further detailed studies are required for understanding  the role of each electron of the 3$d$ orbitals of the Cr$^{2+}$ ions  as well as the  origin of the dual nature, which leads to deeper understandings of the electron correlations in SrCr$_2$As$_2$ and also in the other Cr-based compounds.

% are required to understand the dual nature in SrCr$_2$As$_2$ .      

%    Our magnetic measurements indicate that the Cr ordered moments are itinerant rather than localized.
 %    From the  analysis of  $^{75}$As NMR line width as well as the $M(H)$ of SrCr$_2$As$_2$ single crystal under the two different magnetic field directions $H$ $||$ $ab$ and  $H$ $||$ $c$ in the AFM state, the application of magnetic field parallel to the magnetic easy axis of the $c$ axis was found to make a spin flop, and  the magnetic anisotropy field of the Cr magnetic moments was estimated to be less than 0.1 T.

   \section{Acknowledgments} 
This research was supported by the U.S. Department of Energy, Office of Basic Energy Sciences, Division of Materials Sciences and Engineering. Ames Laboratory is operated for the U.S. Department of Energy by Iowa State University under Contract No.~DE-AC02-07CH11358.

\end{document}